\documentclass[aps,pre,preprint,galleyproof,superscriptaddress,showpacs]{revtex4}%groupedaddress
\usepackage{graphicx,epsf,dsfont,amssymb,verbatim}

\begin{document}
\title{Nonequilibrium statistical operator method in the Renyi
statistics}

\author{B. Markiv}
\email[]{markiv@icmp.lviv.ua}\affiliation{Institute for Condensed
Matter Physics of the National Academy of Sciences of Ukraine,
UA--79011 Lviv, Ukraine}
\author{R. Tokarchuk}%\email[]{}
\affiliation{National university ''Lvivska politechnika'',
UA--79013 Lviv, Ukraine}
\author{P. Kostrobij}%\email[]{}
\affiliation{National university ''Lvivska politechnika'',
UA--79013 Lviv, Ukraine}
\author{M. Tokarchuk}%\email[]{}
\affiliation{Institute for Condensed Matter Physics of the
National Academy of Sciences of Ukraine, UA--79011 Lviv, Ukraine}
\affiliation{National university ''Lvivska politechnika'',
UA--79013 Lviv, Ukraine}
%\date{August 29, 2010}

\begin{abstract}

The generalization of the Zubarev nonequilibrium statistical
operator (NSO) method for the case of Renyi statistics is proposed
when the relevant statistical operator (or distribution function)
is obtained based on the principle of maximum for the Renyi
entropy. The nonequilibrium statistical operator and corresponding
generalized transport equations for the reduced-description
parameters are obtained. A consistent description of kinetic and
hydrodynamic processes in the system of interacting particles is
considered as an example.
\end{abstract}

\maketitle

\section{Introduction}
\label{1}

In investigations of complex self-organizing, fractal structures
and various physical phenomena such as subdiffusion, turbulence,
chemical reactions as well as various economical, social and
biological systems Gibbs distribution function does not provide
agreement with observable phenomena. For these systems the power
distributions are inherent~\citep{1}. They can not be obtained
from the maximum entropy principle for the Boltzmann-Gibbs entropy
underlaying both an equilibrium and nonequilibrium statistical
thermodynamics~\citep{zub1,zub2,3}.

In papers of A.G.~Bashkirov~\citep{4,5,6,7} the use of the
Renyi~\citep{8,9} entropy as statistical entropy for investigation
of complex systems is proposed. It depends on the parameter $q$
($0<q\leq 1$) and at $q=1$ it coincides with Boltzmann-Gibbs
entropy. Based on the maximum entropy principle for the Renyi
entropy the power Renyi distribution was obtained in the case of
equilibrium. At $q=1$ it develops into the Gibbs canonical
distribution. Herewith $\eta=1-q$ is considered as an order
parameter. With the increase of it the statistical Renyi entropy
grows to its maximum which the power Renyi distribution
corresponds to. Moreover the Renyi entropy derivative with respect
to $\eta$ suffers a sudden change at $\eta=0$, that is a sort of
phase transition to more ordered equilibrium state takes place.
The papers of Abe~\citep{12,13}, Arimitsu~\citep{14,15},
Lesche~\citep{16}, Masi~\citep{161} and others were devoted to
axiomatic argumentation and problems of stability of the Renyi
entropy and its linearized form~--- the Havrda-Charvat-Tsallis
entropy~\citep{10,11}. The properties of the Renyi entropy are
discussed in books~\citep{8,17,18} as well. Nowadays the Tsallis
entropy is widely applied in various directions of nonextensive
statistical mechanics~\citep{19,20,21}. The examples are the
phenomena of subdiffusion~\citep{22} and turbulence~\citep{23,24},
the investigations of transport coefficients in gases and
plasma~\citep{25,26} as well as quantum dissipative
systems~\citep{261}. The problems of the construction of
equilibrium thermodynamics in the framework of the Renyi and
Tsallis statistics were discussed
in~\citep{27,28,29,30,31,32,33,34}. The energy
fluctuations~\citep{30}, kinetics of nonequilibrium
plasma~\citep{35}, problems of self-gravitating
systems~\citep{36}, ozone layer~\citep{37}, universality in
non-Debye relaxation~\citep{38} and complex systems~\citep{39,40}
were investigated within the Tsallis statistics.

In the present paper one approach of formulation of an extensive
statistical mechanics of nonequilibrium processes is considered
based on the nonequilibrium statistical operator method by
Zubarev~\citep{zub1,zub2,3} and the maximum entropy principle for
the Renyi entropy. The consistent description of kinetic and
hydrodynamic processes in the system of classical interacting
particles is considered as an example.

\section{Maximum entropy principle for the Renyi entropy
and nonequilibrium statistical operator} \label{2}

Nonequilibrium state of a classical or quantum system of
interacting particles is completely described by the
nonequilibrium statistical operator (nonequilibrium distribution
function) $\varrho(x^N;t)$. The latter satisfies the classical or
quantum Liouville equation

\begin{eqnarray}
\label{math/2} \frac{\partial}{\partial t}\varrho(x^N;t)
+iL_N\varrho(x^N;t)=0.
\end{eqnarray}
 $iL_N$ is the Liouville operator the system of interacting particles
which in classical case has the following form:
\begin{eqnarray}
\label{math/1}
iL_N=\sum_{l=1}^N\frac{\vec{p}_l}{m}\cdot\frac{\partial}{\partial
\vec{r}_{l}}-\frac{1}{2}\sum_{l\neq j=1}^N\frac{\partial}{\partial
\vec{r}_{l}}\Phi(r_{lj})\left(\frac{\partial}{\partial
\vec{p}_l}-\frac{\partial}{\partial \vec{p}_j}\right).
\end{eqnarray}
Here $x_j=\{\vec{p}_j,\vec{r}_{j}\}$ is the phase variables of
$j$-particle, $\Phi(r_{lj})$ is the interaction energy of two
particles, $\vec{p}_j$ the $j$-particle momentum and $m$ its mass,
$r_{lj}=|\vec{r}_{l}-\vec{r}_{j}|$ the distance between pair of
interacting particles.

The function $\varrho(x^N;t)$ is symmetric regarding to inversion
of phase variables of any pair of particles $x_l\leftrightarrow
x_j$ and satisfies the normalization condition
$\int d\Gamma_N\varrho(x^N;t)=1$, $d\Gamma=(dx)^N/N!$,
$dx=d\vec{p}d\vec{r}$.

In order to solve the Liouville equation (\ref{math/2}) we will
use the Zubarev nonequilibrium statistical operator
method~\citep{zub1,zub2,3}. Within its framework we will be
looking for solutions of the equation (\ref{math/2}), which are
independent of the initial conditions. The solutions will depend
on time explicitly only, i.e. through the observable quantities
\begin{equation}
\int d\Gamma_N\hat{P}_n\varrho(x^N;t)=\langle\hat{P}_n\rangle^t
\end{equation}
selected for the reduced description of nonequilibrium states of
the system. In particular, for the description of the hydrodynamic
state of the system of classical interacting particles the
nonequilibrium averaged values of densities of  number of
particles $\langle\hat{n}(\vec{r})\rangle^t$, their momentum
$\langle\hat{\vec{p}}(\vec{r})\rangle^t$ and total energy
$\langle\hat{\varepsilon}(\vec{r})\rangle^t$ can be chosen as the
parameters of the reduced description~\citep{zub1,zub2,3}. They
satisfy the respective conservation laws. In the case of kinetic
description the one-particle $f_{1}(\vec{p},\vec{r};t)$ and
two-particle $f_{2}(\vec{p},\vec{r},\vec{p}',\vec{r}';t)$
distribution functions can serve as the reduced-description
parameters. For the investigation of properties of magnetic and
polar systems the averaged values of densities of magnetic
$\langle\hat{\vec{m}}(\vec{r})\rangle^t$ and dipole
$\langle\hat{\vec{d}}(\vec{r})\rangle^t$ moments can be used
respectively besides the hydrodynamic variables.

When the basic parameters of the reduced description are defined,
the solution $\varrho(x^N;t)$ can be presented by means of the
nonequilibrium statistical operator method in the following
general form~\citep{zub1,zub2,3}:
\begin{eqnarray}
\label{math/26}
\varrho(x^N;t)=\varrho_{rel}(x^N;t)-\int_{-\infty}^te^{\varepsilon(t'-t)}T(t,t')
\left(1-P_{rel}(t')\right)iL_N\varrho_{rel}(x^N;t')dt',
\end{eqnarray}
where
$T(t,t')=\exp_+\Bigl\{-\int_{t'}^t\left(1-P_{rel}(t')\right)iL_Ndt'\Bigr\}$
is the evolution operator containing projection; $\exp_+$ the
ordered exponential. $P_{rel}(t')$ is the generalized
Kawasaki-Gunton projection operator whose structure depends on the
form of the relevant statistical operator (distribution function)
$\varrho_{rel}(x^N;t)$. The latter will be determined using
maximum entropy principle for the Renyi entropy
\begin{equation}
\label{math/3'} S^R(\varrho)=\frac{1}{1-q}\ln\int
d\Gamma_N\varrho^q(t).
\end{equation}
The corresponding functional at fixed parameters of the reduced
description with taking into account the normalization condition
has the form
\begin{equation}
\label{math/3} L^R(\varrho)=\frac{1}{1-q}\ln\int
d\Gamma_N\varrho^q(t)-\alpha\int
d\Gamma_N\varrho(t)-\sum_nF_n(t)\int d\Gamma_N\hat{P}_n\varrho(t),
\end{equation}
$F_n(t)$ are the Lagrange multipliers. Equalizing its functional
derivative to zero %%
%\begin{equation}
%\label{math/4} \frac{\delta L^R(\varrho)}{\delta
%\varrho}=\frac{q}{1-q}\frac{\varrho^{q-1}(t)}{\int
%d\Gamma_N\varrho^q(t)}-\alpha-\sum_nF_n(t)\hat{P}_n=0,
%\end{equation}
%%
we obtain the relevant statistical operator corresponding to the
Renyi entropy maximum:
\begin{equation}
\label{math/14}
\varrho_{rel}(t)=\frac{1}{Z_R}\left(1-\frac{q-1}{q}\sum_nF_n(t)\delta\hat{P}_n\right)
^{\frac{1}{q-1}},
\end{equation}
\begin{equation}
\label{math/15} Z_R(t)=\int
d\Gamma_N\left(1-\frac{q-1}{q}\sum_nF_n(t)\delta\hat{P}_n\right)
^{\frac{1}{q-1}}.
\end{equation}
$Z_R(t)$ is the partition function of the relevant statistical
operator,
$\delta\hat{P}_{n}=\hat{P}_{n}-\langle\hat{P}_n\rangle^t$, and the
parameter $\alpha$ in (\ref{math/3}) is determined by the relation
\begin{equation}
\label{math/6}
\alpha=\frac{q}{1-q}-\sum_nF_n(t)\langle\hat{P}_n\rangle^t.
\end{equation}

The Lagrange multipliers $F_n(t)$ in
(\ref{math/14})-(\ref{math/6}) are defined from the
self-consistency conditions:
\begin{equation}
\label{math/15'}
\langle\hat{P}_n\rangle^t=\langle\hat{P}_n\rangle_{rel}^t,
\end{equation}
$\langle\ldots\rangle^t_{rel}=\int
d\Gamma_N\ldots\varrho_{rel}(x^N;t)$. Since the relevant
statistical operator is now known for the basic set of the
reduced-description parameters, we can obtain the nonequilibrium
statistical operator, establishing the structure of the projection
operator:
\begin{eqnarray}
\label{math/18}&&
P_{rel}(t)\varrho'=\left(\varrho_{rel}(t)-\sum_n\frac{\delta
\varrho_{rel}(t)}{\delta
\langle\hat{P}_n\rangle^t}\langle\hat{P}_n\rangle^t\right)\int
d\Gamma_N \varrho'
\\&&\mbox{}
+\sum_n\frac{\delta \varrho_{rel}(t)}{\delta
\langle\hat{P}_n\rangle^t}\int
d\Gamma_N\hat{P}_n\varrho'.\nonumber
\end{eqnarray}
The variation derivative of the relevant statistical operator in
(\ref{math/18}) can be presented in the form:
\begin{eqnarray}
\label{math/24} \frac{\delta \varrho_{rel}(t)}{\delta
\langle\hat{P}_m\rangle^t}=\varrho_{rel}(t)\delta\left[
\frac{1}{q}\psi^{-1}(t)\left(F_m(t) -\sum_n\frac{\delta
F_n(t)}{\delta\langle\hat{P}_m\rangle^t}\delta\hat{P}_n\right)\right],
\end{eqnarray}
where
\begin{eqnarray}
\label{math/25}
\delta\left[\ldots\right]=\left[\ldots\right]-\langle\left[\ldots\right]\rangle^t_{rel},
\end{eqnarray}
and we use the notation
\begin{eqnarray}
\label{math/22}
\psi(t)=1-\frac{q-1}{q}\sum_nF_n(t)\delta\hat{P}_n.
\end{eqnarray}
The Lagrange multipliers derivatives with regard to the
reduced-description parameters $\delta
F_n(t)/\delta\langle\hat{P}_m\rangle^t$ we calculate in the
following way
\begin{eqnarray}
\label{math/30} \frac{\delta
F_n(t)}{\delta\langle\hat{P}_m\rangle^t}=\left(\frac{\delta\langle\hat{P}_m\rangle^t}{\delta
F_n(t)}\right)^{-1}.
\end{eqnarray}
It can be done in general case. Thus
\begin{eqnarray}
\label{math/30'} \frac{\delta\langle\hat{P}_m\rangle^t}{\delta
F_n(t)}=\frac{\delta\langle\hat{P}_m\rangle^t_{rel}}{\delta
F_n(t)}=\int
d\Gamma_{N}\hat{P}_m\frac{\delta\varrho_{rel}(t)}{\delta F_n(t)}.
\end{eqnarray}
After calculating ${\delta\varrho_{rel}(t)}/{\delta F_n(t)}$ in
the right-hand side of the relation (\ref{math/30'}), we obtain
the set of equations for desired derivatives
\begin{eqnarray}
\label{math/30''} \frac{\delta\langle\hat{P}_m\rangle^t}{\delta
F_n(t)}=\frac{1}{q}\langle\delta\hat{P}_m
\psi^{-1}(t)\rangle^t_{rel}
\sum_l\frac{\delta\langle\hat{P}_l\rangle^t}{\delta
F_n(t)}-\frac{1}{q}\langle\delta\hat{P}_m
\psi^{-1}(t)\delta\hat{P}_n\rangle^t_{rel}.
\end{eqnarray}
Its solution in the matrix form is
\begin{eqnarray}
\label{math/30'''} \frac{\delta\langle\hat{P}\rangle^t}{\delta
F(t)}=-\left[I-\frac{1}{q}\langle\delta\hat{P}
\psi^{-1}(t)\rangle^t_{rel}F\right]^{-1}
\frac{1}{q}\langle\delta\hat{P}
\psi^{-1}(t)\delta\hat{P}\rangle^t_{rel}=f(t),
\end{eqnarray}
where $I$ is the unit matrix and
\begin{eqnarray}
\label{math/30''''} \frac{\delta\langle\hat{P}_m\rangle^t}{\delta
F_n(t)}=\left(\frac{\delta\langle\hat{P}\rangle^t}{\delta
F(t)}\right)_{mn}=f_{mn}(t).
\end{eqnarray}
Thus the functional derivative can be written in the form:
\begin{eqnarray}
\label{math/31} \frac{\delta
\varrho_{rel}(t)}{\delta\langle\hat{P}_m\rangle^t}=\varrho_{rel}(t)\delta
\left[\frac{1}{q} \psi^{-1}(t)\left(F_m(t)
+\sum_nf_{mn}^{-1}(t)\delta\hat{P}_n\right)\right].
\end{eqnarray}
Then the Kawasaki-Gunton projection operator has the following
structure:
\begin{eqnarray}
\label{math/34}\lefteqn{ P_{rel}(t)\varrho'=\varrho_{rel}(t) \int
d\Gamma_N\{\varrho'\} }
\\&&\mbox{}+\sum_m\varrho_{rel}(t)\delta
\left[ \frac{1}{q}\psi^{-1}(t)\left(F_m(t)
+\sum_nf_{mn}^{-1}(t)\delta\hat{P}_n\right)\right]\nonumber
\\&&\mbox{}\times \left(\int
d\Gamma_N\{\hat{P}_m\varrho'\}-\langle\hat{P}_m\rangle^t\int
d\Gamma_N\{\varrho'\}\right).\nonumber
\end{eqnarray}
It is further necessary to explore an action of the operators
$P_{rel}(t)iL_N$ on the relevant statistical operator. Since
\begin{eqnarray}
\label{math/35} iL_N\varrho_{rel}(t)=-\varrho_{rel}(t)
\frac{1}{q}\psi^{-1}(t)\sum_nF_n(t)\dot{\hat{P}}_n
=A(t)\varrho_{rel}(t),
\end{eqnarray}
then
\begin{eqnarray}
\label{math/36}\lefteqn{
P_{rel}(t)iL_N\varrho_{rel}(t)=P(t)A(t)\varrho_{rel}(t)= \int
d\Gamma_N\{A(t)\varrho_{rel}(t)\} }
\\&&\mbox{}+\sum_m\varrho_{rel}(t)\delta
\left[ \frac{1}{q}\psi^{-1}(t)\left(F_m(t)
+\sum_nf_{mn}^{-1}(t)\delta\hat{P}_n\right)\right]\nonumber
\\&&\mbox{}\times
\left(\int
d\Gamma_N\{\hat{P}_mA(t)\varrho_{rel}(t)\}-\langle\hat{P}_m\rangle^t\int
d \Gamma_N\{A(t)\varrho_{rel}(t)\}\right),\nonumber
\end{eqnarray}
 where
\begin{eqnarray}
\label{math/37}\int
d\Gamma_N\{\hat{P}_mA(t)\varrho_{rel}(t)\}-\langle\hat{P}_m\rangle^t
\int d\Gamma_N\{A(t)\varrho_{rel}(t)\} =\langle\delta\hat{P}_m
A(t)\rangle^t_{rel}.
\end{eqnarray}
Thus
$P_{rel}(t)iL_N\varrho_{rel}(t)=P_{rel}(t)A(t)\varrho_{rel}(t)=(P(t)A(t))\varrho_{rel}(t)$,
where $P(t)$ is the projection operator which now acts on dynamic
variables:
\begin{eqnarray}
\label{math/39} \lefteqn{P(t)\ldots=\langle\ldots\rangle_{rel}^t}
\\&&\mbox{}+\sum_m \delta \left[ \frac{1}{q}\psi^{-1}(t)\left(F_m(t)
+\sum_nf_{mn}^{-1}(t)\delta\hat{P}_n\right)\right]
\langle\ldots\delta\hat{P}_m\rangle_{rel}^t.\nonumber
\end{eqnarray}
So far as
\begin{eqnarray}
\label{math/40} A(t)=-\frac{1}{q} \psi^{-1}(t)\sum_nF_n(t)
\dot{\hat{P}}_n,
\end{eqnarray}
we have
\begin{eqnarray}
\label{math/41}\lefteqn{ P(t)A(t)=-\frac{1}{q}\sum_nF_n(t) \langle
\psi^{-1}(t)\dot{\hat{P}}_n\rangle_{rel}^t}
\\&&\mbox{}+\sum_m\delta
\left[ \frac{1}{q}\psi^{-1}(t)\left(F_m(t)
+\sum_{l}f_{ml}^{-1}(t)\delta\hat{P}_l\right)\right]\nonumber
\\&&\mbox{}\times
\left\langle\Bigl[-\frac{1}{q} \psi^{-1}(t)\sum_nF_n(t)
\dot{\hat{P}}_n-\right.\nonumber
\\&&\mbox{}\left.\frac{1}{q}\sum_nF_n(t)
\langle \psi^{-1}(t)\dot{\hat{P}}_n\rangle_{rel}\Bigr] (\hat{P}_m
-\langle\hat{P}_m\rangle_{rel})\right\rangle_{rel}^t.\nonumber
\end{eqnarray}
Considering (\ref{math/35})-(\ref{math/41}) we can present
$(1-P_{rel}(t))iL_N\varrho_{rel}(t)$ in the form:
\begin{eqnarray}
\label{math/42} \lefteqn{(1-P_{rel}(t))iL_N\varrho_{rel}(t)=}
\\&&\mbox{}=(1-P(t))iL_N\varrho_{rel}(t)
=-\sum_nI_n(t)F_n(t)\varrho_{rel}(t),\nonumber
\end{eqnarray}
where
\begin{eqnarray}
\label{math/43} I_n(t)=(1-P(t))
\frac{1}{q}\psi^{-1}(t)\dot{\hat{P}}_{n}
\end{eqnarray}
are the generalized flows. Taking into account (\ref{math/42}) we
can now write down an explicit expression for the nonequilibrium
statistical operator
\begin{eqnarray}
\label{math/269} \lefteqn{\varrho(x^N;t)=\varrho_{rel}(x^N;t)}
\\&&\mbox{}+\sum_n\int_{-\infty}^te^{\varepsilon(t'-t)}T(t,t')
 I_n(t')F_n(t')\varrho_{rel}(x^N;t')dt'.\nonumber
\end{eqnarray}

It allows us to obtain the generalized transport equations for the
reduced-description parameters. They can be presented in the form:
\begin{eqnarray}
\label{math/44}\frac{\partial}{\partial
t}\langle\hat{P}_m\rangle^t=\langle\dot{\hat{P}}_m\rangle^t_{rel}
+\sum_n\int_{-\infty}^te^{\varepsilon(t'-t)}\varphi_{mn}(t,t')F_n(t')dt',
\end{eqnarray}
with the generalized transport kernels (memory functions)
\begin{eqnarray}
\label{math/45}\varphi_{mn}(t,t')= \int
d\Gamma_N\{\dot{\hat{P}}_mT(t,t')I_n(t')\varrho_{rel}(t')\}
\end{eqnarray}
which describe the dissipative processes in the system.

\section{Generalized transport equations for a consistent
description of kinetics and hydrodynamics in the Renyi statistics}
\label{3}

For a consistent description of kinetic and hydrodynamic processes
in classical (or quantum) systems of $N$ particles interacting in
the volume $V$ the nonequilibrium one-particle distribution
function $f_1(x;t)=\langle\hat{n}_1(x)\rangle^t$ and averaged
value of potential energy of interaction
$\varepsilon_{int}(\vec{r};t)=\langle\hat{\varepsilon}_{int}(\vec{r})\rangle^t$
are the basic parameters of the reduced description~\citep{TOK}.
The last is defined through the two-particle distribution function
$f_2(x,x';t)=\langle\hat{n}_2(x,x')\rangle^t$. Here $\hat{n}_1(x)$
and $\hat{n}_2(x,x')$ are the phase densities of the microscopic
distribution of particles, and
$\hat{\varepsilon}_{int}(\vec{r})=\frac{1}{2}\int d\vec{p}\int
d\vec{p'}\int d\vec{r'}\Phi(|\vec{r}-\vec{r'}|)\hat{n}_2(x,x')$.
In this case according to (\ref{math/14}) we obtain the relevant
statistical operator
\begin{equation}
\label{math/140}
\varrho_{rel}(t)=\frac{1}{Z_R}\left(1-\frac{q-1}{q}
\Bigl\{ \int d\vec{r}\beta (\vec{r};t)\delta\hat{\varepsilon}_{int}(\vec{r};t)\\
+\int dx a(x;t)\delta
\hat{n}_1(x;t)\Bigr\}\right)^{\frac{1}{q-1}},\nonumber
\end{equation}
where
\begin{equation}\label{math/150}
Z_R(t)=\int d\Gamma_N\left(1-\frac{q-1}{q}\Bigl\{ \int d\vec{r}
\beta (\vec{r};t)\delta\hat{\varepsilon}_{int}(\vec{r};t)\\
+\int dx a(x;t)\delta
\hat{n}_1(x;t)\Bigr\}\right)^{\frac{1}{q-1}}\nonumber
\end{equation}
is the partition function. The parameters $\beta (\vec{r};t)$ and
$a(x;t)$ are defined from the self-consistency conditions:
\begin{equation}\label{math/1500}
\langle\hat{\varepsilon}_{int}(\vec{r})\rangle^t
=\langle\hat{\varepsilon}_{int}(\vec{r})\rangle_{rel}^t, \qquad
\langle\hat{n}_1(x)\rangle^t=\langle\hat{n}_1(x)\rangle_{rel}^t.
\end{equation}

According to (\ref{math/26}) and taking into consideration
(\ref{math/140}) we write down the nonequilibrium statistical
operator of a consistent description of kinetic and hydrodynamic
processes in the system:
\begin{eqnarray}\label{math/260}
\lefteqn{\varrho(x^N;t)=\varrho_{rel}(x^N;t)+\int_{-\infty}^te^{\varepsilon(t'-t)}T(t,t')}
\\&&\mbox{}\times\left(\int d\vec{r}'\beta
(\vec{r}';t')I_{\varepsilon}^{int}(\vec{r}';t')+ \int dx'
a(x';t')I_{n}(x';t')\right)\varrho_{rel}(x^N;t')dt'.\nonumber
\end{eqnarray}
Here
\begin{equation}\label{math/2601}
I_{n}(x';t')=(1-P(t'))\frac{1}{q}\psi^{-1}(t')iL_N \hat{n}_1(x')
\end{equation}
\begin{equation}\label{math/2602}
I_{\varepsilon}^{int}(\vec{r}';t')=(1-P(t'))\frac{1}{q}\psi^{-1}(t')iL_N
\hat{\varepsilon}_{int}(\vec{r}')
\end{equation}
are the generalized flows. With the help of the nonequilibrium
statistical operator obtained one can derive the generalized
transport equations for the basic set of the reduced-description
parameters according to (\ref{math/44}):
\begin{eqnarray}
\label{math/440}\lefteqn{\frac{\partial}{\partial
t}\langle\hat{n}_1(x)\rangle^t=\frac{1}{q}\int dx' \Phi_{n
\dot{n}}(x,x';t)a(x';t)}
\\&&\mbox{}+\frac{1}{q}\int d\vec{r}\Phi_{n
\dot{\varepsilon}}(x,\vec{r};t) \beta (\vec{r};t)+\int
dx'\int_{-\infty}^te^{\varepsilon(t'-t)}\varphi_{nn}(x,x';t,t')a(x';t')dt'\nonumber
\\&&\mbox{}+\int d\vec{r}'\int_{-\infty}^te^{\varepsilon(t'-t)} \varphi_{n
\varepsilon}(x,\vec{r}';t,t')\beta (\vec{r}';t')dt',\nonumber
\end{eqnarray}
\begin{eqnarray}
\label{math/450}\lefteqn{\frac{\partial}{\partial
t}\langle\hat{\varepsilon}_{int}(\vec{r})\rangle^t
=\frac{1}{q}\int dx' \Phi_{\varepsilon
\dot{n}}(\vec{r},x';t)a(x';t)}
\\&&\mbox{}+\frac{1}{q}\int
d\vec{r}'\Phi_{\varepsilon \dot{\varepsilon}}(\vec{r},\vec{r}';t)
\beta (\vec{r}';t)+\int dx'\int_{-\infty}^te^{\varepsilon(t'-t)}
\varphi_{\varepsilon n}(\vec{r},x';t,t')a(x';t')dt' \nonumber
\\&&\mbox{}+\int d\vec{r}'\int_{-\infty}^te^{\varepsilon(t'-t)}
\varphi_{\varepsilon \varepsilon}(\vec{r},\vec{r}';t,t')\beta
(\vec{r}';t')dt'.\nonumber
\end{eqnarray}
The functions
\begin{equation}
\Phi_{n \dot{n}}(x,x';t)=\int d\Gamma_N\hat{n}_1(x)
\psi^{-1}(t)iL_N \hat{n}_1(x')\varrho_{rel}(x^N;t),
\end{equation}
\begin{equation}
\Phi_{n \dot{\varepsilon}}(x,\vec{r}';t)=\int
d\Gamma_N\hat{n}_1(x) \psi^{-1}(t)iL_N
\hat{\varepsilon}_{int}(\vec{r}')\varrho_{rel}(x^N;t),
\end{equation}
\begin{equation}
\Phi_{\varepsilon \dot{n}}(\vec{r},x';t)=\int
d\Gamma_N\hat{\varepsilon}_{int}(\vec{r}) \psi^{-1}(t)iL_N
\hat{n}_1(x')\varrho_{rel}(x^N;t),
\end{equation}
\begin{equation}
\Phi_{\varepsilon \dot{\varepsilon}}(\vec{r},\vec{r}';t)=\int
d\Gamma_N\hat{\varepsilon}_{int}(\vec{r}) \psi^{-1}(t)iL_N
\hat{\varepsilon}_{int}(\vec{r}')\varrho_{rel}(x^N;t),
\end{equation}
are the time correlation one obtained by means of the relevant
distribution and which contain the function
\begin{equation}
\psi(t)=1-\frac{q-1}{q}\left( \int d\vec{r}\beta (\vec{r};t)
\delta\hat{\varepsilon}_{int}(\vec{r};t)\\
+\int dx a(x;t)\delta \hat{n}_1(x;t)\right),
\end{equation}
At $q=1$ $\psi(t)=1$, we have the transition to the relevant Gibbs
distribution when the nonequilibrium one-particle distribution
function $f_1(x;t)=\langle\hat{n}_1(x)\rangle^t$ and averaged
value of potential energy of interaction
$\varepsilon_{int}(\vec{r};t)=\langle\hat{\varepsilon}_{int}(\vec{r})\rangle^t$
are the parameters of the reduced description~\citep{TOK}. The
generalized transport kernels $\varphi_{nn}(x,x';t,t')$,
$\varphi_{n \varepsilon}(x,\vec{r}';t,t')$, $\varphi_{\varepsilon
n}(\vec{r},x';t,t')$, $\varphi_{\varepsilon
\varepsilon}(\vec{r},\vec{r}';t,t')$ have the structure of
(\ref{math/45}) with the corresponding flows (\ref{math/2601}),
(\ref{math/2602}).

For the case when the nonequilibrium one-particle distribution
function $f_1(x;t)=\langle\hat{n}_1(x)\rangle^t$ is the single
parameter of the reduced description (the contribution of
nonequilibrium averaged potential energy of interaction is much
smaller then the kinetic energy, e.g. the case of rare gases) the
set of transport equations (\ref{math/440})-(\ref{math/450})
reduces to the generalized kinetic equation:
\begin{eqnarray}
\label{math/441}\lefteqn{\frac{\partial}{\partial
t}\langle\hat{n}_1(x)\rangle^t=\frac{1}{q}\int dx' \Phi_{n
\dot{n}}(x,x';t)a(x';t)} \\&&\mbox{} +\int
dx'\int_{-\infty}^te^{\varepsilon(t'-t)}\varphi_{nn}(x,x';t,t')a(x';t')dt'.\nonumber
\end{eqnarray}
At $q=1$ it transforms into the kinetic equation of~\citep{TOK}
with the transport kernel calculated using relevant distribution
function
$\varrho_{rel}(t)=\prod_{j=1}^{N}\frac{f_{1}(x_{j};t)}{e}$. In
this case at $q=1$ within the NSO method~\citep{zub1,zub2,3} the
Liouville equation (\ref{math/2}) should be solved with the
boundary condition
\begin{eqnarray} \label{math/2222}
\frac{\partial}{\partial t}\varrho(x^N;t)
+iL_N\varrho(x^N;t)=-\varepsilon
\left(\varrho(x^N;t)-\prod_{j=1}^{N}\frac{f_{1}(x_{j};t)}{e}\right).
\end{eqnarray}
It corresponds to the Bogolyubov hypothesis of weakening of the
correlations between particles. Integration of this equation with
respect to phase variables $\int d\Gamma_{N-1}$, $\int
d\Gamma_{N-2}$....$\int d\Gamma_{N-s}$ leads to the BBGKY
hierarchy for the nonequilibrium particle distribution functions.
The question about the BBGKY hierarchy when in the boundary
condition (\ref{math/2222}) $\varrho_{rel}(t)$ is equal to
(\ref{math/140}) is interesting. With $q=1$ $\varrho_{rel}(t)$
transforms into the Gibbs form and we obtain the BBGKY hierarchy
with the modified boundary condition~\citep{zub2,3,TOK} which
takes into account many-particle correlations. It allows one to
obtain in the pair collision approximation the revised Enskog
theory and the Enskog-Landau kinetic equations for the neutral and
charged hard sphere system,
respectively~\citep{zub5,zub6,zub2,3,TOK}.

\section{Conclusions}
\label{4}

For the nonequilibrium system of interacting particles within the
framework of the Zubarev NSO method we obtained the nonequilibrium
statistical operator $\varrho(t)$. It satisfies the Liouville
equation with the boundary condition describing the relaxation of
the NSO to the relevant statistical operator $\varrho_{rel}(t)$.
The latter is constructed based on the maximum entropy principle
for the Renyi entropy at fixed values of the reduced-description
parameters $\langle\hat{P}_n\rangle^t$ with taking into account
the normalization condition. By means of the NSO the generalized
transport equations for the parameters of the reduced description
$\langle\hat{P}_n\rangle^t$ are obtained with regard to
Kawasaki-Gunton and Mori projection. Such an approach is applied
to a consistent description of kinetic and hydrodynamic processes
in the system of classical interacting particles. As a result both
the nonequilibrium statistical operator and the generalized
transport equations are obtained, when the nonequilibrium
one-particle distribution function
$f_1(x;t)=\langle\hat{n}_1(x)\rangle^t$ along with the
nonequilibrium averaged value of the potential energy of
interaction $\varepsilon_{int}(\vec{r};t)$ are selected as the
reduced-description parameters. At $q=1$ the known results based
on the Gibbs statistics are reproduced. Naturally, an interesting
question about the investigation of time correlation functions and
transport coefficients based on the NSO method within the Renyi
statistics arises.

\bibliographystyle{elsarticle-harv}
\bibliography{<your-bib-database>}



\end{document}